# Generation and propagation characterization of a vortex beam through an electro-optical crystal-based electrically controlled flat plate

Yuting Fan[a], Enxu Zhu[a] and Chaoying Zhao[a,b*]

[a]Hangzhou Dianzi University, College of Science, Zhejiang, 310018, China

[b]Shanxi University, State Key Laboratory of Quantum Optics and Quantum Optics Devices, Taiyuan, 030006, China

*Corresponding author: zchy49@163.com

**Abstract:** With the increasing demand for potential applications in almost all fields in modern optics, the generation of vortex beams has attracted significant interest. Based on a flat plate made of electro-optical crystals, we propose an electrically controlling method to generate vortex beams assisted by the Pockels effect. Compared with traditional methods, our method allows flexibly tuning the wavelength and topological charge of the generated optical vortex beams. We simulate the propagation of optical beams transmitted from the flat plate and investigate the orbital-angular-momentum(OAM)-mode spectra of the transmitted optical beams. The OAM-mode spectra are in agreement well with the simulation results. The proposed method improves the flexibility of the generation of optical vortex beams, may gain more significant potential applications in optical communication and optical manipulation.

**OCIS:** Optical vortices(050.4865), Singular optics(260.6042), Interference(260.3160)

## 1. Introduction

Optical vortex beams with helical wavefront associated with azimuthally dependent phase $e^{il\varphi}$, where integer $l$ is termed topological charge (TC) and $\varphi$ is the azimuthal angle, carry orbital angular momentum (OAM) $l\hbar$ per photon [1], which in principle can constitute an infinite dimension of Hilbert space while the spin angular momentum (SAM) only corresponds to two eigenstates, namely left-hand circular polarization and right-hand circular polarization. To date, increasing applications based on optical vortexes have been proposed both in classical optics and quantum optics, such as optical communication [2], optical manipulation [3, 4], imaging and microscopy [5], and quantum information processing [6].

Since the OAM of light was discovered [1], lots of optical components have been proposed to convert non-OAM light into optical vortexes. Spiral phase plates (SPPs), thin transparent plates with changed thickness with respect to the azimuthal angle, add an azimuthally changed phase on the passed-through optical beams, resulting in generating optical vortex beams with TCs [7,8]. However, each fabricated SPP can only output optical vortex beams with a designed TC, and the deviation between the wavelength of incident optical beams and the designed wavelength would decrease the purity of the OAM mode [9]. The inhomogeneous and anisotropic plate with different local directions of the optical axis, named q-plate, can convert circularly polarized non-OAM incident optical beams into optical vortex beams with handedness-inverted circular polarization and TC equaling to $\pm 2q$ [10], where $q$ is the TC of the q-plate. However, q-plates made of liquid crystal also lack reconfigurability even though the conversion efficiency and operating wavelength can be electrically controlled [11,12]. Employing programmable spatial light modulators (SLMs) enables one to generate optical vortex beams more flexibly [13]. Nevertheless, the SLM composed of a series of discrete optical components is usually bulky. Although metasurfaces with dimension-designed nanoblocks determining the local optical phase [14,15] and ring-shape whispering-gallery-mode (WGM) microresonators [16] whose inner wall is embedded angular grating structures bring optical-vortex emitters and lasers into an ultracompact area, the solution of flexibly tuning the wavelength and TC of generated optical vortex beams still remains an attractive research topic.

Dynamically tunable optical vortex beams would release more advanced applications. For example, unbounded wavelengths and OAM modes make wavelength-division multiplexing (WDM) or time-division multiplexing (TDM) combined with OAM mode-division multiplexing (MDM) possible, which can significantly increase the information capacity of the optical communication system [17]. SPPs combined with a Fabry-Perot filter [18] or using the combination of different vector modes of various optical fibers [19, 20] leads to the tunability of the wavelength and TC. Tunable lasers supporting Hermite-Gaussian modes which can be converted into OAM modes through external converters give rise to a relatively wide range of tunability of the TC [21,22]. Recently, Zhang et al. reported a non-Hermitian symmetry-breaking WGM microring resonator on the InGaAsP multiple quantum well platform, where the controllable OAM-mode switch was achieved [23]. In addition, Ji et al. utilized a photocurrent made of two-dimensional material $WTe_2$ to experimentally detect the TC of the received optical vortex beam [24], which indicates integrated direct-electrical-readout OAM detectors have the potential to be realized, and implies the requirement of electrically tunable vortex-beam sources.

The Pockels effect describes the linear change in refractive index with respect to the applied electric field, and it is often used in the design of various electro-optical elements [25,26]. Here, based on the Pockels effect, we suggest a flat thin plate made of electro-optical crystal (potassium dihydrogen phosphate [KDP] crystal as an example in this paper) with external electric fields to control the local refractive index resulting in modulating the phase of the incident light with respect to the azimuthal angle, finally converting linearly polarized non-OAM optical beams into optical vortex beams with desired TCs. In our configuration, independently controlling the electric field applied on each phase-modulated region would change the TC of the output optical vortex beam with an arbitrary wavelength of the incident light.

## 2. Concept and model

Fig.1 shows the schematic of an SPP and a flat plate with regions modulated by the external electric field for the generation of optical vortex beams.

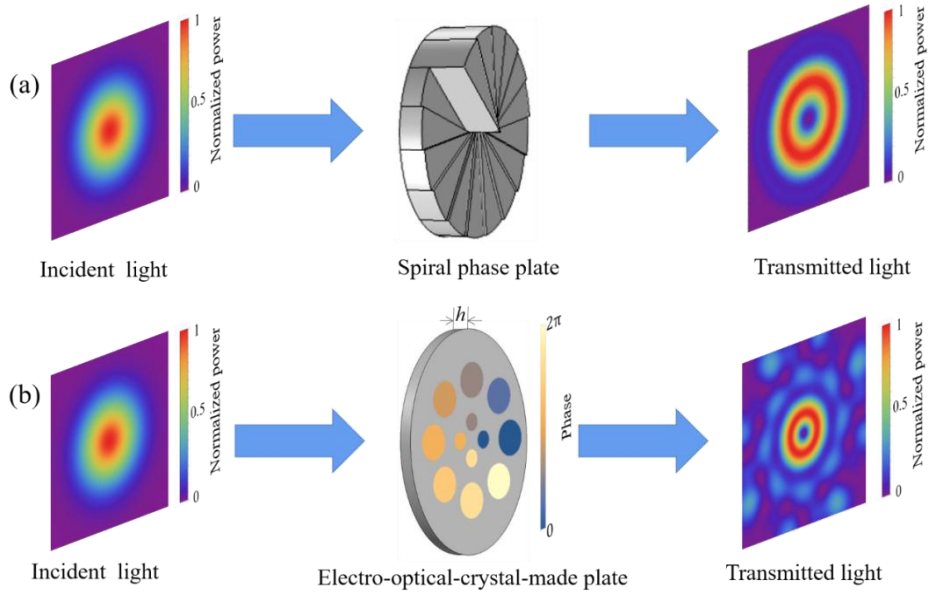

**Fig. 1** The generation of optical vortex beams (a) through a traditional SPP. The incident light has a Gaussian intensity distribution. (b) through a flat plate made of electro-optical crystals, where the local refractive index is electrically controlled by a micro-electrode plate array so that the phase of the incident light is partially modulated.

As Fig.1(a) shows, the multi-level SPP is designed with azimuthally increased thickness to modulate the phase of the incident optical beam, giving rise to a helical wavefront of the output light whose TC is determined by the thickness difference between the steps of the SPP. However, the value of the thickness difference is not tunable once the SPP is fabricated. In addition to varying the propagating distance in the medium, controlling the refractive index also leads to the modulation of the optical phase. We propose a flat plate made of electro-optical crystal so that the refractive index can be electrically controlled with the help of the Pockels effect[27]. Considering azimuthally changed continuous electric field with a singularity in the center is hard to construct, we use independently controlled micro-electrode plate array[28] made of transparent conducting materials (such as tin-doped indium oxide [29]) to modulate the local phase of the incident non-OAM optical beams (such as the Gaussian beam) as shown in Fig.1(b).

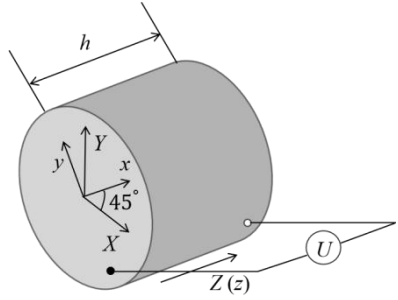

**Fig.2** Schematic of a micro-electrode plate. The $XYZ$ coordinate is the principal-axis coordinate of the KDP crystal without electric field modulation. The optical axis of the KDP crystal alongs the $Z$ direction. After appling an external electric field, due to the electromechanical light modulation, the principal axis rotation about the $Z$-axis 45° with the help of the Pockels effect[27].

The electromechanical light modulation in uniaxial crystals have been discussed in Ref. [30]. As an illustrative example, we consider the flat plate is made of the KDP crystal so that the optical axis is parallel to the $z$-axis and perpendicular to the entrance face (see Fig.2). Here, we consider the incident light propagating along the $z$-axis is linearly polarized in the $y$-axis. Our research is based on the fact that the voltage of each electrode plate can be controlled independently due to the electromechanical light modulation. The electrode plate array is parallel to the entrance face and generates electric fields in the $z$-direction applied to discrete modulation regions. Thus, the local refractive index of each region can be independently tuned by controlling voltages of the electrode plates based on the Pockels effect. The relation between the refractive index $n_y$ for $y$-direction linearly polarized light and the magnitude of the applied electric field $E_z$ can be written as [27]

$$n_y = n_o + \frac{1}{2}n_o^3 r_{63} E_z, \tag{1}$$

where $n_o$ is the refractive index for the ordinary light in the absence of an external electric field, and $r_{63}$ is the electro-optical coefficient of the KDP crystal. After the light transmits from the plate, the phase difference $\Delta\theta$ between the modulated region and the unmodulated region is

determined by

$$\Delta\theta = \theta - \theta_0 = k_0 n_y h - k_0 n_0 h, \quad (2)$$

where $\theta$ is the designed phase of the modulated region, $\theta_0$ is the relative phase of the unmodulated region, it doesn't matter what the value is, and $k_0 = 2\pi/\lambda$ is the wavenumber, $\lambda$ is the wavelength of the incident light, and $h$ is the thickness of the flat plate. Thus, the refractive index of the modulated region should satisfy

$$n_y = \frac{\theta - \theta_0}{k_0 h} + n_0. \quad (3)$$

By substituting Eq. (3) into Eq. (1), the corresponding voltage of the modulated region is determined by

$$U = E_z h = \frac{2(\theta - \theta_0)}{n_0^3 r_{63} k_0}. \quad (4)$$

By utilizing wavelength-tunable lasers as the incident-light sources, independently tuning the voltage of each electrode plate to modulate each local phase of the incident light would lead to generating wavelength- and OAM-tunable optical vortex beams. The OAM mode can be determined by the phase difference $\Delta\theta$. Without loss of generality, we assume $\theta_0 = \pi/4$. For a KDP crystal: the refractive index $n_o = 1.514$, the electro-optical coefficient $r_{63} = 10.5 \times 10^{-12}$ m/V, the incident light wavelength $\lambda = 0.5461$ μm (green light). In numerical simulation, we add a backward voltage $U = -3.747$ kV to attains $\theta = 0$, and we add a forward voltage $U = 26.23$ kV to attains $\theta = 2\pi$. The magnitudes of the voltages is in agreement with that of the references[31,32]. Employing electro-optical crystals with stronger nonlinearity allows significantly decreasing the operating voltage, for example, the lithium niobite (LiNbO3) crystal has been successfully used in electro-optic frequency comb generators, which the peak voltage equals to 10V [33].

### 3. Simulation results and mode-spectrum analysis

In this section, we consider the local refractive index of each electric-field-applied region is finely controlled so that the phase of the transmitted light is modulated. We use the Gaussian beam as the incident light, and simulate the propagations of the transmitted optical beam in the free space by the diffraction theory [34].

Fig. 3 shows the transverse field distribution of the transmitted optical beam propagating in free space at different distances. The first column shows the modulated regions of the incident light denoted with black circles. In Fig. 3(a), we tune the voltage of the electrode plate array to achieve a $2\pi$ phase shift to construct optical beams with TC=1. In the beginning, the transverse pattern is broken up. As the beam propagates further, the transverse optical field recombines to form a C-shaped pattern with a rotation characteristic at $z=1\text{ m}$. Moreover, we find that the pattern becomes stable (the shape does not change when the propagation distance still increases) and presents a spiral tail, which is identical to the interference pattern formed by a vortex beam with TC=1 and a co-propagating Gaussian beam [15]. We note that there are fragmentary patterns with insignificant power in the output beam. We attribute it to the discontinuity of the modulation regions. In Fig. 3(b), we invert the phase shift to construct optical beams with TC=−1. The corresponding optical field distributions are similar to the ones in Fig. 3(a) except that the rotation direction is inverse. When we construct a two-$2\pi$ phase shift (Fig. 3(c) or (d)), the very beginning transverse pattern still has a broken shape, and gradually forms a stable interference-like pattern with two spiral tails, which indicates the output beam is superposed by the TC=0 mode and TC=2 (or TC=−2) mode. We also simulated the case that only a $\pi$ phase shift is achieved (Fig. 3(e)). Interestingly, the stable transverse pattern also has a spiral tail. However, the solid light spot in the center is larger than the one in the $2\pi$-phase-shift case, which indicates the relative weight of the non-OAM mode is relatively large.

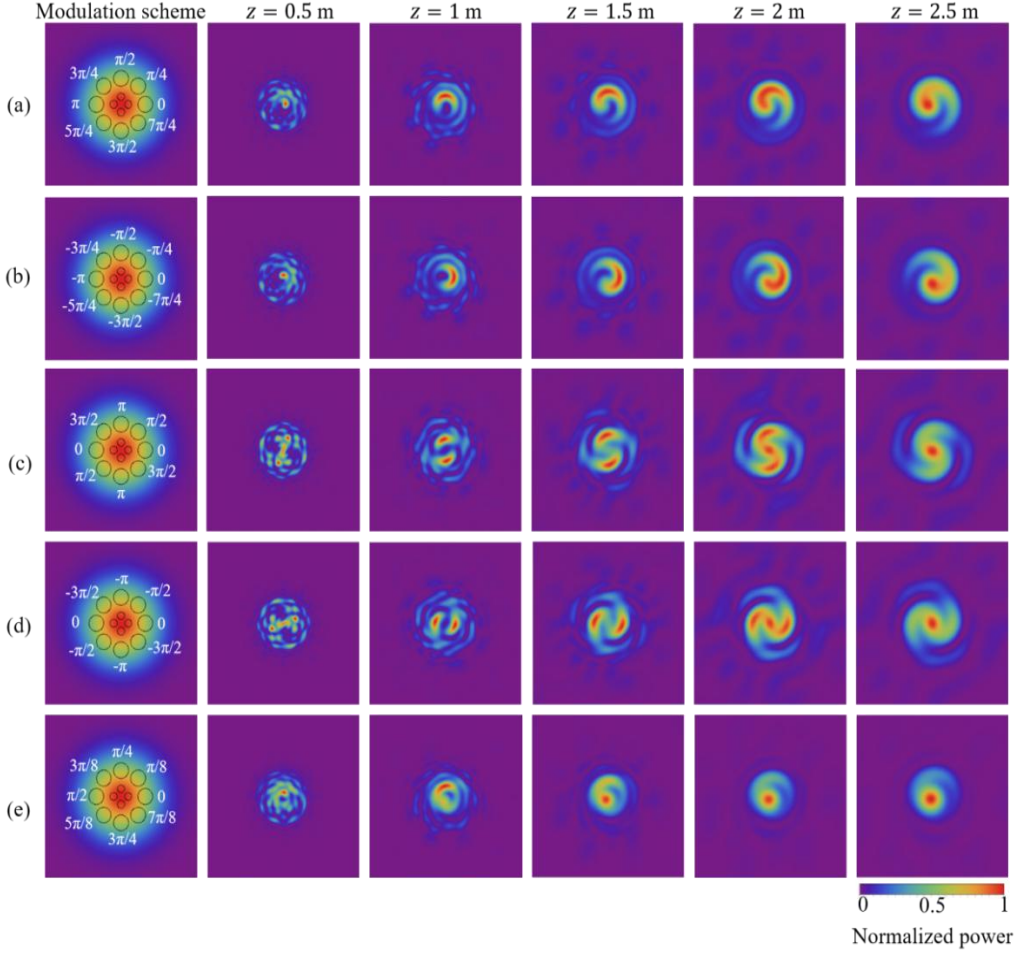

**Fig. 3** Propagation of the transmitted light in free space for the case that the whole incident light passes through the plate. The black circles in the first column denote the modulation regions of the incident light which has a Gaussian intensity distribution, and the white values represent the azimuthally modulated optical phase whose corresponding applied voltages can be calculated by Eq. (4). The corresponding optical phase of the unmodulated region is set by $\pi/4$. (a) $2\pi$ phase shift; (b) inverse $2\pi$ phase shift; (c) two-$2\pi$ phase shift; (d) inverse two-$2\pi$ phase shift; (e) $\pi$ phase shift.

Now we consider the region where no external electric field applies is blocked, i.e., only the modulated regions of the incident light can transmit from the flat plate. The transverse-field distributions of different phase shifts are shown in Fig. 4. The first column shows the distribution of the optical field in the transmission surface and the phase modulation schemes. Fig. 4(a) shows the case that no external electric field applies, which is equivalent to the pinhole diffraction. With light propagating in the free space, a solid light spot formed in the center, which is similar to the Airy spot. In Fig. 4 (b) (or (c)), we construct a $2\pi$ phase shift. The complex transverse field gradually evolves into a stable state with a doughnut-shaped pattern indicating it carries TC = 1 (or TC =− 1 ). We note that the propagation distance for stabilizing the transverse pattern is shorter than the one in Fig. 3. When we contrast a two-$2\pi$ phase shift to generate the optical vortex with TC = 2 (TC =− 2), the stable transverse pattern also shows a doughnut-like shape. However, the intensity of the "doughnut" is slightly inhomogeneous, and four relatively high-power fields around the "doughnut" are present. We attribute this phenomenon to that there are only two values of 0 and π of the optical phase in the inner modulated region. Generating optical vortex with higher TC requires more modulated regions in the azimuthal direction. For the situation that only a π phase shift is created, the stable distribution of the transverse optical field is composed by a solid light spot and a C-shape pattern with fragmentary optical fields around the center light spot.

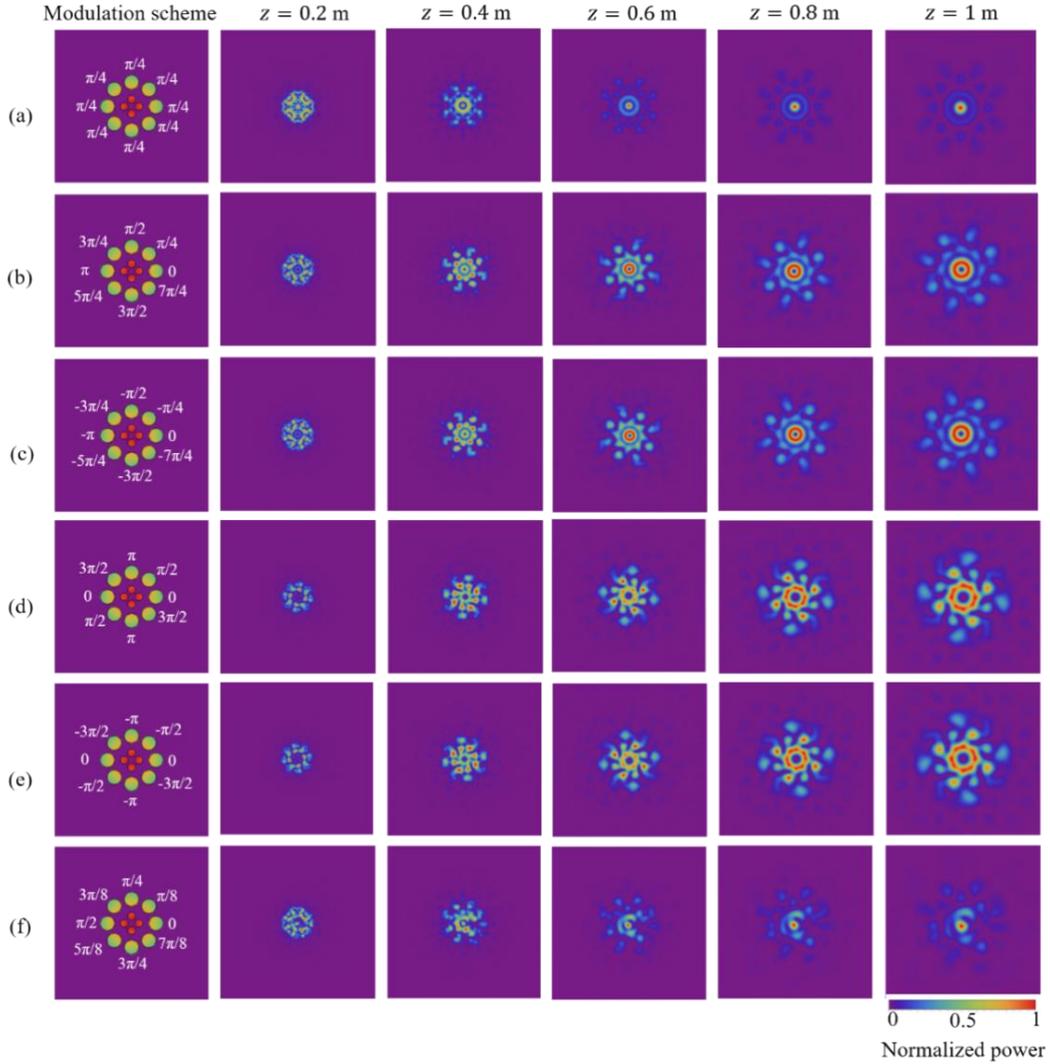

**Fig. 4** Propagation of the transmitted light in a free space when the incident light only pass through the modulated regions of the plate. The first column shows the modulation schemes for (a)without external electric field, π/4 phase (b)2π phase shift, (c)inverse 2π phase shift, (d)two-2π phase shift, (e)inverse two-2π phase shift, and (f)π phase shift.

To know the relative weight of each OAM mode of the generated optical beams, we now analyze the mode spectrum. The OAM modes ($A_l \sim e^{-il\theta}$) are orthogonal to each other, i.e.,

$$\int A_l^* A_{l'} r dr d\theta = 0, \text{ for } l \neq l'. \tag{5}$$

The OAM modes are complete in a Hilbert space so that any optical beam in the space can be decomposed by the orthogonal OAM-mode basis, and hence we have

$$A(r,\theta,z) = \frac{1}{\sqrt{2\pi}}\sum_{l=-\infty}^{\infty} a_l(r,z)e^{-il\theta}, \qquad (6)$$

where $A(r,\theta,z)$ is the complex amplitude of the optical field. The superposition factor $a_l(r,z)$ satisfies

$$a_l(r,z) = \frac{1}{\sqrt{2\pi}}\int_0^{2\pi} A(r,\theta,z)e^{-il\theta}d\theta. \qquad (7)$$

By integrating $|a_l(r,z)|^2$ with respect to $r$, we define

$$C_l = \int_0^\infty |a_l(r,z)|^2 r dr \qquad (8)$$

representing the power of the $l$-th mode [35]. Hence, the relative power of the $l$-th mode is given by

$$P_l = \frac{C_l}{\sum_{m=-\infty}^{\infty} C_m}. \qquad (9)$$

Fig. 5 shows the OAM-mode spectrum of the output beam with different phase modulation schemes for the case that both modulated and unmodulated regions of the incident light can transmit from the flat plate.

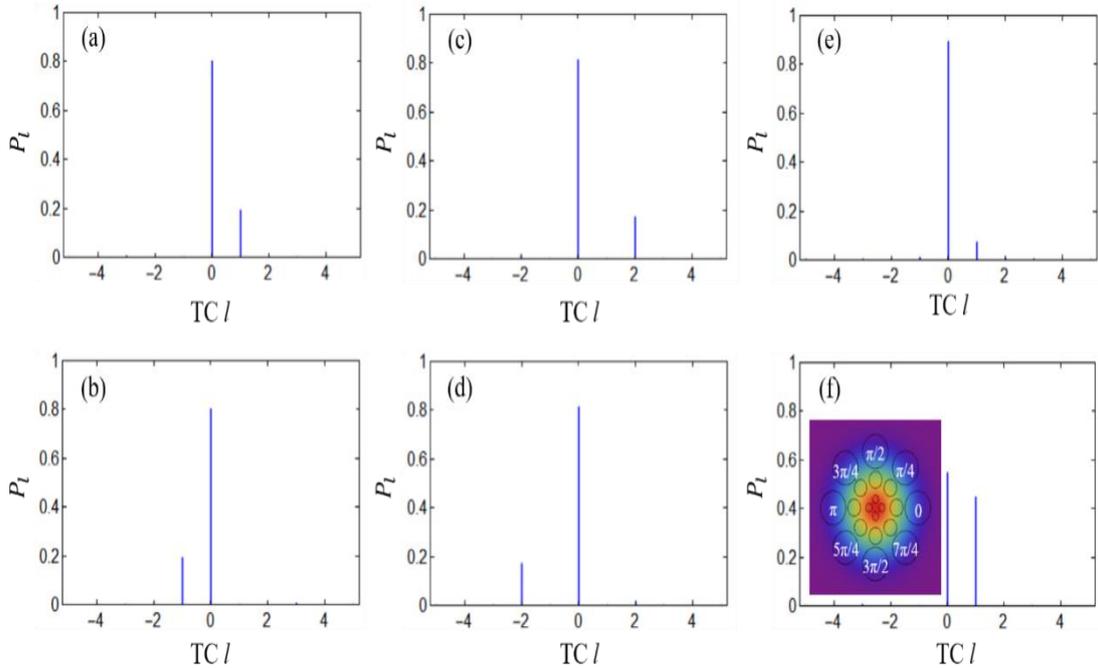

**Fig. 5** OAM-mode spectra with different modulation schemes for the case that the whole incident light passes through the plate. The modulation schemes of (a-e) are the same as in Fig. 3(a-e), respectively. (f) The mode spectrum with the insert showing the modulation scheme, in which a $2\pi$ phase shift is also achieved but the area of modulated regions is more than the one in (a).

As we expect, the mode with TC = 1 occurs in Fig. 5(a) for the $2\pi$-phase-shift modulation scheme identical to the first row in Fig. 3. However, the spectrum also shows the presence of the non-OAM mode with a high relative weight. This is because the unmodulated region of the light carrying non-OAM mode also transmits from the flat plate. The simultaneous presence of the non-OAM mode and the vortex mode explains the interference-like pattern in Fig. 3. Increasing the modulation region leads to the decrease in the relative weight of the non-OAM mode as shown in Fig. 5(f). For the $\pi$-phase-shift modulation scheme, the mode with TC = 1 also occurs, but the relative weight is lower than the one in Fig. 5(a).

Fig. 6 shows the OAM-mode spectrum of the output beam with different phase modulation schemes for the case that only modulated regions of the incident light can transmit from the flat plate. The results indicate the output beams are with high purity of OAM modes for the $2\pi$-phase-shift modulation scheme. We note that the mode with TC = 2 (TC =− 2) occurs with a small relative weight for the (inverse) two-$2\pi$-phase-shift modulation scheme. This can be accounted for that there are only two values in the $2\pi$ period of the optical phase in the inner modulated region. When only a half phase period is achieved (Fig. 6(e)), the output beam simultaneously carries the non-OAM component and vortex component with TC = 1 as well as other components with an insignificant relative weight.

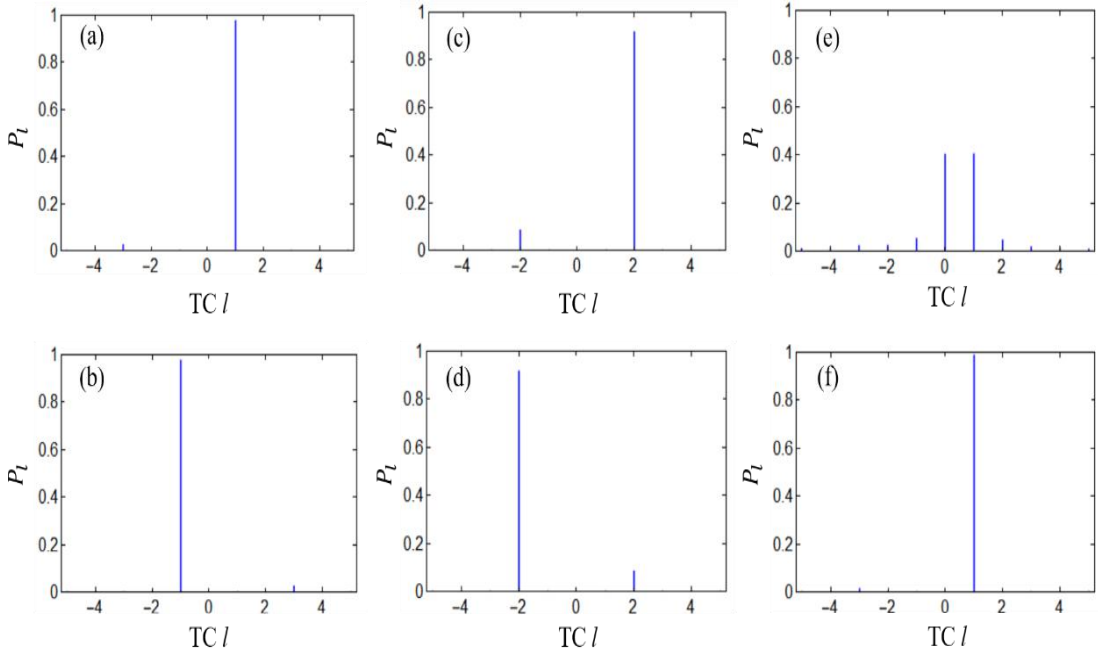

**Fig. 6** OAM-mode spectra with different modulation schemes for the case that only the modulated regions of the incident light pass through the plate. The modulation schemes of (a-f) are the same as in Fig. 5(a-f), respectively.

## 4. Summary

We propose a flat plate made of electro-optical crystal to generate optical vortex beams. The TC of the generated beam can be controlled by the external electric field according to the wavelength. The transverse patterns of the output optical beams are investigated when the unmodulated region of the incident light can pass through or cannot pass through the flat plate. The interference-like patterns indicate the output light contains both OAM component and non-OAM component when the whole incident light can pass through the flat plate. The purity of the OAM mode can be improved by blocking the unmodulated region. We also investigate the OAM-mode spectra, which show a great agreement with the property of the transverse patterns. This work may gain more significant potential applications in various wavelength- and OAM-tunable optical-vortex lasers and emitters.

**Acknowledgments**

This work was supported by the State Key Laboratory of Quantum Optics and Quantum Optics Devices, Shanxi University, Shanxi, China (KF202004).

**REFERENCES**

1. L. Allen, M. W. Beijersbergen, R. J. C. Spreeuw, and J. P. Woerdman, "Orbital angular momentum of light and transformation of Laguerre Gaussian Laser modes*,"* Phys. Rev. A, **45**(11),8185-8189(1992).

2. Z. Xie, S. Gao, T. Lei, S. Feng, Y. Zhang, F. Li, J. Zhang, Z. Li, and X. Yuan, "Integrated (de)multiplexer for orbital angular momentum fiber communication," Photon. Res., **6**(7),743-749(2018).

3. M. Gecevičius, R. Drevinskas, M. Beresna, and P. G. Kazansky, "Single beam optical vortex tweezers with tunable orbital angular momentum," Appl. Phys. Lett., **104**(23), 231110(2014).

4. Y. Li, L. M. Zhou, and N. Zhao, "Anomalous motion of a particle levitated by Laguerre–Gaussian beams," Opt. Lett., **46**(1),106-109(2021).

5. Y. Kozawa, D. Matsunaga, and S. Sato, "Superresolution imaging via superoscillation focusing of a radially polarized beam," Optica, **5**(2),86-92(2018).

6. T. Stav, A. Faerman, E. Maguid, D. Oren, V. Kleiner, E. Hasman, and M. Segev, "Quantum entanglement of the spin and orbital angular momentum of photons using metamaterials," Science, **361**(6407),1101-1104(2018).

7. M. W. Beijersbergen, R. P. C. Coerwinkel, M. Kristensen, and J. P. Woerdman,

"Helical-wavefront laser beams produced with a spiral phaseplate," Opt. Commun., 1994.

8. K. Sueda, G. Miyaji, N. Miyanaga, and M. Nakatsuka, "Laguerre-Gaussian beam generated with a multilevel spiral phase plate for high intensity laser pulses," Opt. Exp., **12**(15),3548-3553(2004).

9. S. N. Khonina, V. V. Podlipnov, S. V. Karpeev, A. V. Ustinov, S. G. Volotovsky, and S. V. Ganchevskaya, "Spectral control of the orbital angular momentum of a laser beam based on 3D properties of spiral phase plates fabricated for an infrared wavelength," Opt. Exp., **28**(12),18407-18417(2020).

10. L. Marrucci, C. Manzo, and D. Paparo, "Optical Spin-to-Orbital Angular Momentum Conversion in Inhomogeneous Anisotropic Media," Phys. Rev. Lett., **96**(16),163905(2006).

11. B. Piccirillo, V. D'Ambrosio, S. Slussarenko, L. Marrucci, and E. Santamato, "Photon spin-to-orbital angular momentum conversion via an electrically tunable q-plate," Appl. Phys. Lett., **97**,241104(2010).

12. E. Brasselet, "Tunable High-Resolution Macroscopic Self-Engineered Geometric Phase Optical Elements," Phys. Rev. Lett., **121**(3),033901(2018).

13. A. Forbes, A. Dudley, and M. McLaren, "Creation and detection of optical modes with spatial light modulators," Adv. in Opt. Photon., **8**(2),200-227(2016).

14. M. I. Shalaev, J. Sun, A. Tsukernik, A. Pandey, K. Nikolskiy, and N. M. Litchinitser, "High-Efficiency All-Dielectric Metasurfaces for Ultracompact Beam Manipulation in Transmission Mode," Nano Letters, **15**(9),6261-6266(2015).

15. N. Yu, P. Genevet, M. A. Kats, F. Aieta, J. P. Tetienne, F. Capasso, and Z. Gaburro, "Light Propagation with Phase Discontinuities: Generalized Laws of Reflection and Refraction," Science, **334**(6054),333-337(2011).

16. X. Cai, J. Wang, M. J. Strain, B. Johnson-Morris, J. Zhu, M. Sorel, J. L. O'Brien, M.G. Thompson, and S. Yu, "Integrated Compact Optical Vortex Beam Emitters," Science, **338**(6105),363-366(2012).

17. A. Wang, L. Zhu, J. Liu, C. Du, Q. Mo, and J. Wang, "Demonstration of hybrid orbital angular momentum multiplexing and time-division multiplexing passive optical network," Opt. Exp., **23**(23),29457-29466(2015).

18. V. S. Lyubopytov, A. P. Porfirev, S. O. Gurbatov, S. Paul, M. F. Schumann, J. Cesar, M.

Malekizandi, M. T. Haidar, M. Wegener, A. Chipouline, and F. Küppers, "Simultaneous wavelength and orbital angular momentum demultiplexing using tunable MEMS-based Fabry-Perot filter," Opt. Exp., **25**(9),9634-9646(2017).

19. W. Zhang, K. Wei, L. Huang, D. Mao, B. Jiang, F. Gao, G. Zhang, T. Mei, and J. Zhao, "Optical vortex generation with wavelength tunability based on an acoustically-induced fiber grating," Opt. Exp., **24**(17),19278-19285(2016).

20. S. Yao, G. Ren, Y. Shen, Y. Jiang, B. Zhu, and S. Jian, "Tunable Orbital Angular Momentum Generation Using All-Fiber Fused Coupler," IEEE Photon. Tech. Lett., **30**(1),99-102(2018).

21. Y. Shen, Y. Meng, X. Fu, and M. Gong, "Wavelength-tunable Hermite-Gaussian modes and an orbital-angular-momentum-tunable vortex beam in a dual-off-axis pumped Yb:CALGO laser," Opt. Lett., **43**(2),291-294(2018).

22. S. Wang, S. l. Zhang, P. Li, M. H. Hao, H. M. Yang, J. Xie, G. Y. Feng, and S. H. Zhou, "Generation of wavelength- and OAM-tunable vortex beam at low threshold," Opt. Exp., **26**(14),18164-18170(2018).

23. Z. Zhang, X. Qiao, B. Midya, K. Liu, J. Sun, T. Wu, W. Liu, R. Agarwal, J. M. Jornet, S. Longhi, N.M. Litchinitser, and L. Feng, "Tunable topological charge vortex microlaser," Science, **368**(6492),760-763(2020).

24. Z. Ji, W. Liu, S. Krylyuk, X. Fan, Z. Zhang, A. Pan, L. Feng, A. Davydov, and R. Agarwal, "Photocurrent detection of the orbital angular momentum of light," Science, **368**(6492),763-767(2020).

25. M. Thomaschewski, V. A. Zenin, C. Wolff, and S. I. Bozhevolnyi, "Plasmonic monolithic lithium niobate directional coupler switches," Nat.Commun., **11**(1), 748(2020).

26. K. Alexander, J. P. George, J. Verbist, K. Neyts, B. Kuyken, D. Van Thourhout, and J. Beeckman, "Nanophotonic Pockels modulators on a silicon nitride platform," Nat. Commun., **9**(1),3444(2018).

27. R. W. Boyd, Nonlinear Optics, Third Edition. 2008: Academic Press.

28. M. Hourmand, A. A. D. Sarhan, and M. Sayuti, "Micro-electrode fabrication processes for micro-EDM drilling and milling: a state-of-the-art review," Int. J. Adv. Manu. Tech., **91**(1),1023-1056(2017).

29. G. U. Kulkarni, S. Kiruthika, R. Gupta, and K. D. M. Rao, "Towards low cost materials and

methods for transparent electrodes," Curr. Opin. Chem. Eng., **8**,60-68(2015).

30. W. Zhu and W. She, "Electrically controlling spin and orbital angular momentum of a focused light beam in a uniaxial crystal," Opt. Exp., **20**(23),25876-25883(2012).

31. S. N. Khonina, V. V. Podlipnov, and S. G. Volotovskiĭ, "Study of the electro-optical transformation of linearly polarized Bessel beams propagating along the optic axis of an anisotropic DKDP crystal," J. Opt. Tech., **85**(7),388-395(2018).

32. H. Chu, Y. Li, and S. Zhao, "Improving deuterated potassium dihydrogen phosphate's electro-optical Q-switched characteristics by adding a pair of auxiliary electrodes," Appl. Opt., **50**(3), 360-365(2011).

33. M. Zhang, B. Buscaino, C. Wang, A. Shams-Ansari, C. Reimer, R. Zhu, J.M. Kahn, and M. Lončar, "Broadband electro-optic frequency comb generation in a lithium niobate microring resonator," Nature, **568**(7752),373-377(2019).

34. G. M. Lao, Z. H. Zhang, and D. M. Zhao, "Propagation of the power-exponent-phase vortex beam in paraxial ABCD system," Opt. Exp., **24**(16),18082-18094(2016).

35. G. Molina-Terriza, J. P. Torres, and L. Torner, "Management of the Angular Momentum of Light: Preparation of Photons in Multidimensional Vector States of Angular Momentum," Phys. Rev. Lett., **88**(1),013601(2001).